\documentclass[twocolumn,prb]{revtex4}
\usepackage{amsfonts}
\usepackage[T1]{fontenc}
\usepackage{amsmath,amsbsy,amssymb,graphicx}
\usepackage{times}

\let\mathbf=\boldsymbol

\begin{document}

\title{{\Large Photo-Induced Topological Phase Transition and a Single
Dirac-Cone State}\\
{\Large in Silicene }}
\author{Motohiko Ezawa}
\affiliation{Department of Applied Physics, University of Tokyo, Hongo 7-3-1, 113-8656,
Japan }

\begin{abstract}
Silicene (a monolayer of silicon atoms) is a two-dimensional topological
insulator (TI), which undergoes a topological phase transition to a band
insulator under external electric field $E_{z}$. We investigate a
photo-induced topological phase transition from a TI to another TI by
changing its topological class by irradiating circular polarized light at
fixed $E_{z}$. The band structure is modified by photon dressing with a new
dispersion, where the topological property is altered. By increasing the
intensity of light at $E_{z}=0$, a photo-induced quantum Hall insulator is
realized. Its edge modes are anisotropic chiral, in which the velocities of
up and down spins are different. At $E_{z}>E_{\text{cr}}$ with a certain
critical field $E_{\text{cr}}$, a photo-induced spin-polarized quantum Hall
insulator emerges. This is a new state of matter, possessing one Chern
number and one half spin-Chern numbers. We newly discovered a single
Dirac-cone state along a phase boundary. A distinctive hallmark of the state
is that one of the two Dirac valleys is closed and the other open.
\end{abstract}

\maketitle


\address{{\normalsize Department of Applied Physics, University of Tokyo, Hongo
7-3-1, 113-8656, Japan }}

\textbf{Introduction:} Topological insulator (TI) is a distinctive state of
matter indexed by topological numbers, and characterized by an insulating
gap in the bulk and topologically protected gapless edges\cite{Hasan,Qi}.
The topological classification has been applied to static systems\cite%
{Schnyder}, but recently extended to time-periodic systems\cite%
{YMQ,Oka09L,Inoue,Kitagawa01B,Lindner,Dora}. A powerful method to drive
quantum systems periodically is to apply electromagnetic radiation to them.
It can rearrange the band structure and change material properties by photon
dressing. TIs may be obtained from a semimetal\cite{Oka09L} and from a band
insulator\cite{Lindner} (BI) in this way. Topological band structures may
well be engineered by application of coherent laser beam in graphene and
semiconductors. In this paper we propose a new type of topological phase
transition in silicene, that is a photo-induced transition from a TI to
another TI by changing its topological class.

Silicene, being synthesized\cite{GLayPRL,Kawai,Takamura} only recently, is
gifted with enormously rich physics\cite%
{LiuPRL,EzawaNJP,Falko,EzawaAQHE,EzawaDiamag,EzawaOpti}. Silicene consists
of a honeycomb lattice of silicon atoms with buckled sublattices made of A
sites and B sites. The states near the Fermi energy are $\pi $ orbitals
residing near the K and K' points at opposite corners of the hexagonal
Brillouin zone. The low-energy dynamics in the K and K' valleys is described
by the Dirac theory as in graphene. However, Dirac electrons are massive due
to a relatively large spin-orbit (SO) coupling $\lambda_{\text{SO}}=3.9$meV
in silicene. It is remarkable that the mass can be controlled\cite%
{EzawaNJP,Falko} by applying the electric field $E_{z}$ perpendicular to the
silicene sheet.

Silicene is a quantum spin Hall insulator\cite{LiuPRL} (QSHI), which is a
particular type of TI. It undergoes a topological phase transition\cite%
{EzawaNJP,Falko} to a BI as $|E_{z}|$ increases and crosses the critical
field $E_{\text{cr}}$. We investigate photo-induced topological phase
transition. Under the off-resonance coherent laser beam, Berry curvatures in
the momentum space, originating from the SO coupling, are modified in the
photon-dressed bands so that the occupied electronic states change
topological properties\cite{YMQ}. The phase diagram has a remarkably rich
structure, where there are three distinct topological phases in addition to
one trivial phase.

We show that, by applying strong circular polarized light with frequency $%
\Omega $ at fixed $E_{z}$, silicene is transformed from a QSHI or a BI into
a photo-induced spin-polarized quantum Hall insulator (PS-QHI) and
eventually into a photo-induced quantum Hall insulator (P-QHI). Here, PS-QHI
is a new state of matter indexed by one Chern and one half spin-Chern
numbers. On the other hand, the edge modes of P-QHI are anisotropic chiral,
where the velocities of up and down spins are different. Furthermore, there
appear spin polarized metal (SPM) and spin valley-polarized metal\cite%
{EzawaAQHE} (SVPM) on the crossing points of the two phase boundaries. They
have different spin configurations. A particularly intriguing state appears
along a phase boundary, which has only one closed gap with a linear
dispersion. We call it the single Dirac-cone (SDC) state. It is utterly a
new state as far as we are aware of. The electric field breaks inversion
symmetry, while the light breaks time-reversal symmetry. When they are both
broken, the gap can be different at K and K' points. We comment that the
Nielsen-Ninomiya theorem\cite{Nielsen}, which states that all massless Dirac
cones must come in pairs, is not applicable to the SDC state since the
chiral symmetry is explicitly broken by the mass term.

\textbf{Tight binding model}: The silicene system is described by the
four-band second-nearest-neighbor tight binding model\cite{EzawaAQHE}, 
\begin{align}
H& =-t\sum_{\left\langle i,j\right\rangle \alpha }c_{i\alpha }^{\dagger
}c_{j\alpha }+i\frac{\lambda _{\text{SO}}}{3\sqrt{3}}\sum_{\left\langle
\!\left\langle i,j\right\rangle \!\right\rangle \alpha \beta }\nu
_{ij}c_{i\alpha }^{\dagger }\sigma _{\alpha \beta }^{z}c_{j\beta }  \notag \\
& +i\lambda _{\text{R1}}(E_{z})\sum_{\left\langle i,j\right\rangle \alpha
\beta }c_{i\alpha }^{\dagger }(\mathbf{\sigma }\times \hat{\mathbf{d}}%
_{ij})_{\alpha \beta }^{z}c_{j\beta }  \notag \\
& -i\frac{2}{3}\lambda _{\text{R2}}\sum_{\left\langle \!\left\langle
i,j\right\rangle \!\right\rangle \alpha \beta }\mu _{i}c_{i\alpha }^{\dagger
}(\mathbf{\sigma }\times \hat{\mathbf{d}}_{ij})_{\alpha \beta }^{z}c_{j\beta
}  \notag \\
& -\ell \sum_{i\alpha }\mu _{i}E_{z}c_{i\alpha }^{\dagger }c_{i\alpha },
\label{BasicHamil}
\end{align}%
where $c_{i\alpha }^{\dagger }$ creates an electron with spin polarization $%
\alpha $ at site $i$, and $\left\langle i,j\right\rangle /\left\langle
\!\left\langle i,j\right\rangle \!\right\rangle $ run over all the
nearest/next-nearest neighbor hopping sites. We explain each term. (i) The
first term represents the usual nearest-neighbor hopping with the transfer
energy $t=1.6$eV. (ii) The second term\cite{KaneMele} represents the
effective SO coupling with $\lambda _{\text{SO}}=3.9$meV, where $\mathbf{%
\sigma }=(\sigma _{x},\sigma _{y},\sigma _{z})$ is the Pauli matrix of spin,
with $\nu _{ij}=+1$ if the next-nearest-neighboring hopping is anticlockwise
and $\nu _{ij}=-1$ if it is clockwise with respect to the positive $z$ axis.
(iii) The third term represents the first Rashba SO coupling associated with
the nearest neighbor hopping, which is induced by external electric field.
It satisfies $\lambda _{\text{R1}}(0)=0$ and becomes of the order of $10\mu $%
eV at the critical electric field $E_{\text{c}}=\lambda _{\text{SO}}/\ell =17
$meV\r{A}$^{-1}$\cite{EzawaAQHE}. (iv) The forth term\cite{LiuPRB}
represents the second Rashba SO coupling with $\lambda _{\text{R2}}=0.7$meV
associated with the next-nearest neighbor hopping term, where $\mu _{i}=\pm 1
$ for the A (B) site, and $\hat{\mathbf{d}}_{ij}=\mathbf{d}_{ij}/\left\vert 
\mathbf{d}_{ij}\right\vert $ with the vector $\mathbf{d}_{ij}$ connecting
two sites $i$ and $j$ in the same sublattice. (v) The fifth term\cite%
{EzawaAQHE} is the staggered sublattice potential term. Due to the buckled
structure the two sublattice planes are separated by a distance, which we
denote by $2\ell $ with $\ell =0.23$\r{A}. It generates a staggered
sublattice potential $\varpropto 2\ell E_{z}$ between silicon atoms at A
sites and B sites in electric field $E_{z}$.

\textbf{Low-energy Dirac theory:} We analyze the physics of electrons near
the Fermi energy, which is described by Dirac electrons near the K and K'
points. We also call them the K$_{\eta }$ points with $\eta =\pm $. The
effective Dirac Hamiltonian in the momentum space reads\cite{EzawaAQHE}%
\begin{align}
H_{\eta }=& \hbar v_{\text{F}}\left( \eta k_{x}\tau _{x}+k_{y}\tau
_{y}\right) +\lambda _{\text{SO}}\sigma _{z}\eta \tau _{z}-\ell E_{z}\tau
_{z}  \notag \\
& +a\eta \tau _{z}\lambda _{\text{R2}}\left( k_{y}\sigma _{x}-k_{x}\sigma
_{y}\right)  \notag \\
& +\lambda _{\text{R1}}\left( E_{z}\right) (\eta \tau _{x}\sigma _{y}-\tau
_{y}\sigma _{x})/2,  \label{DiracHamilSilic}
\end{align}%
where $\sigma _{a}$ and $\tau _{a}$ are the Pauli matrices of the spin and
the sublattice pseudospin, respectively. The first term arises from the
nearest-neighbor hopping, where $v_{\text{F}}=\frac{\sqrt{3}}{2}at=5.5\times
10^{5}$m/s is the Fermi velocity with the lattice constant $a=3.86$\AA .
There is no recognizable effect from the term $\lambda _{\text{R1}}\left(
E_{z}\right) $ as far as we have numerically checked. Although we include
all terms in numerical calculations, in order to simplify the formulas and
to make the physical picture clear, we set $\lambda _{\text{R1}}(E_{z})=0$
in all analytic formulas.

There are four bands in the energy spectrum of $H_{\eta }$. The band gap is
located at the K and K' points, and given by $2|\Delta (E_{z})|$, where\cite%
{EzawaAQHE}%
\begin{equation}
\Delta (E_{z})=\eta s_{z}\lambda _{\text{SO}}-\ell E_{z},  \label{DiagoE}
\end{equation}%
with the spin $s_{z}=\pm 1$. It is a good quantum number at the K and K'
points. The spin $s_{z}$ is an almost good quantum number even away from the
K and K' points because $\lambda _{\text{R2}}$ is a small quantity.

As $|E_{z}|$ increases, the gap decreases linearly, and closes at the
critical point $|E_{z}|=E_{\text{cr}}$ with%
\begin{equation}
E_{\text{cr}}=\lambda _{\text{SO}}/\ell =17\text{meV/\AA },  \label{StepA}
\end{equation}%
and then increases linearly. Silicene is a semimetal due to gapless modes at 
$|E_{z}|=E_{\text{cr}}$, while it is an insulator for $|E_{z}|\neq E_{\text{%
cr}}$. It is to be noted that the change of the gap is suppressed by the
screening effect due to the polarization of the A and B sublattices\cite%
{Falko}. Even if the effect is taken into account, however, the gap changes
linearly as a function of the external field. Hence, the present results
remain true provided the external field is renormalized appropriately.

\textbf{Photo-induced topological insulator:} We consider a beam of
circularly polarized light irradiated onto the silicene sheet. The
corresponding electromagnetic potential is given by%
\begin{equation}
\mathbf{A}(t)=(A\sin \Omega t,A\cos \Omega t),
\end{equation}%
where $\Omega $ is the frequency of light with $\Omega >0$ for the right
circulation and $\Omega <0$ for the left circulation. The light intensity is
characterized by the dimensionless number $\mathcal{A}=eAa/\hbar $, where $%
\mathcal{A}$ is typically less than $1$ for intensity of lasers and pulses
available in the frequency regime of our interests. The gauge potential
satisfies the time-periodicity, $\mathbf{A}(t+T)=\mathbf{A}(t)$, with $%
T=2\pi /|\Omega |$. The electromagnetic potential is introduced into the
Hamiltonian (\ref{DiracHamilSilic}) by way of the minimal substitution, that
is, replacing the momentum $\hbar k_{i}$ with the covariant momentum $%
P_{i}\equiv \hbar k_{i}+eA_{i}$.

A question arises whether a topological classification is possible in
non-equilibrium situation, that is, when the Hamiltonian has an explicit
time dependence. The answer is yes, provided the potential is time-periodic.
A convenient method is to use the Floquet theory\cite%
{Oka09L,Inoue,Kitagawa01B,Lindner,Dora}. The topological classification is
possible with the aid of the static effective Hamiltonian appropriately
constructed.

We summarize the result of the floquet theory. When the light frequency is
off-resonant for any electron transitions, light does not directly excite
electrons and instead effectively modifies the electron band structures
through virtual photon absorption processes. Such an off-resonant condition
is satisfied for the frequency $\hbar |\Omega |\gg t$ in our model with $\pi 
$ bands. The influence of such off-resonant light is summarized\cite%
{Kitagawa01B} in the static effective Hamiltonian defined by $\Delta H_{%
\text{eff}}=(i\hbar /T)\log U$, where $U=\mathcal{T}\exp [-i/\hbar
\int_{0}^{T}H\left( t\right) dt]$ is the time evolution operator with $%
\mathcal{T}$\ the time-ordering operator. In the limit of $\mathcal{A}\ll 1$%
, $\Delta H_{\text{eff}}$\ is particularly simple near the Dirac points: $%
\Delta H_{\text{eff}}=(\hbar \Omega )^{-1}\left[ H_{-1},H_{+1}\right]
+O\left( \mathcal{A}^{4}\right) $, where $H_{\pm 1}$ is the Fourier
component of Hamiltonian, that is, $H_{\pm 1}=\frac{1}{T}\int_{0}^{T}H\left(
T\right) e^{\pm it|\Omega |}dt$. The modification of the Hamiltonian due to
the time-periodic perturbation is understood as the sum of two second-order
processes, where electrons absorb and then emit a photon, and vice versa.

\begin{figure}[t]
\centerline{\includegraphics[width=0.5\textwidth]{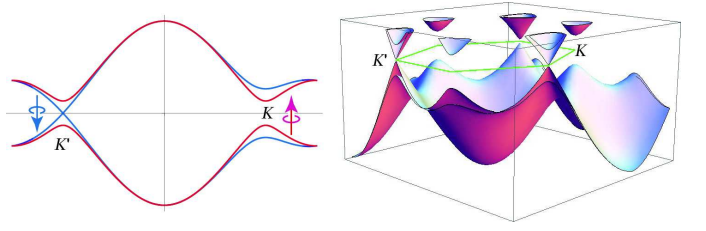}}
\caption{(Color online) The band structure of a silicene in the SDC state.
The gap is open at the K point with a parabolic dispersion but closed at the
K' point with a linear dispersion.}
\label{FigSDBand3D}
\end{figure}

By explicitly calculating the commutation, we have the effective
Hamiltonian, $H_{\text{eff}}^{\eta }=H_{\eta }+\Delta H_{\text{eff}}$, with%
\begin{eqnarray}
\Delta H_{\text{eff}} &=&-\frac{\mathcal{A}^{2}}{\hbar \Omega }[\left( \hbar
v_{\text{F}}\right) ^{2}\eta \tau _{z}-\left( a\lambda _{\text{R2}}\right)
^{2}\sigma _{z}  \notag \\
&&\qquad -a\lambda _{\text{R2}}\hbar v_{\text{F}}(\eta \tau _{x}\sigma
_{y}-\tau _{y}\sigma _{x})].  \label{DeltaEffec}
\end{eqnarray}%
It modifies the band structure. The modification is remarkable, which we
demonstrate based on analytic formulas by neglecting the second Rashba terms
($\varpropto \lambda _{\text{R2}}$) since $\lambda _{\text{R2}}$ is a small
constant. The gap is given by $2|m_{\text{D}}|$ with the Dirac mass,%
\begin{equation}
m_{\text{D}}=\eta s_{z}\lambda _{\text{SO}}-\ell E_{z}-\eta \hbar v_{\text{F}%
}^{2}\mathcal{A}^{2}\Omega ^{-1}.  \label{DiracMass}
\end{equation}%
Hence we can control the Dirac mass by applying electric field $E_{z}$
and/or coherent laser beam $\varpropto \mathcal{A}^{2}/\Omega $. It is to be
emphasized that the band gaps at the K and K' points can be made different
in general. We can make one Dirac cone gapless and the other Dirac cone
gapped. This is the SDC state. The realization of different band gaps at the
two valleys is an entirely new phenomenon. In Fig.\ref{FigSDBand3D} we have
illustrated the band structure of the SDC state calculated based on the
tight-binding Hamiltonian (\ref{BasicHamil}) together with the inclusion of
the Haldane term (\ref{HaldaneTerm}) we discuss later.

\textbf{Spin-Chern number:} Each topological phase is indexed by the
topological quantum numbers, which are the Chern number $\mathcal{C}$ and
the $\mathbb{Z}_{2}$ index. If the spin $s_{z}$ is a good quantum number,
the $\mathbb{Z}_{2}$ index is identical to the spin-Chern number $\mathcal{C}%
_{s}$ modulo $2$. They are defined by $\mathcal{C}=\mathcal{C}_{+}+\mathcal{C%
}_{-}$ and $\mathcal{C}_{s}=\frac{1}{2}(\mathcal{C}_{+}-\mathcal{C}_{-})$,
where $\mathcal{C}_{\pm }$ is the summation of the Berry curvature in the
momentum space over all occupied states of electrons with $s_{z}=\pm 1$.
They are well defined even if the spin is not a good quantum number\cite%
{Prodan09B,Sheng}. In the present model the spin is not a good quantum
number because of spin mixing due to the Rashba couplings $\lambda _{\text{R1%
}}$ and $\lambda _{\text{R2}}$. A convenient way is to calculate them in the
system without the Rashba couplings and then adiabatically switching on
these couplings to recover the present system\cite{Prodan09B,Sheng}.

\begin{figure}[t]
\centerline{\includegraphics[width=0.3\textwidth]{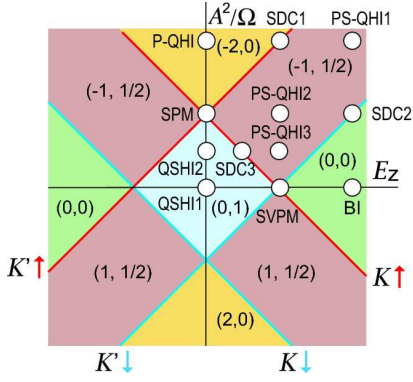}}
\caption{(Color online) Phase diagram in the $(E_{z},\mathcal{A}^{2}/\Omega
) $ plane. A circle shows a point where the energy spectrum is calculated
and shown in Fig.\protect\ref{FigPhotoRibbon}. Heavy lines represent phase
boundaries indexed by K$_{\protect\eta }$ and $s_{z}=\uparrow \downarrow $.
There appear a SDC state along the line, which is characterized by a single
gapless Dirac cone at the K$_{\protect\eta }$ point with spin $s_{z}$. The
topological charges $(C,C_{s})$ are also indicated.}
\label{FigPhaseEA}
\end{figure}

When we set $\lambda _{\text{R1}}=\lambda _{\text{R2}}=0$, the Hamiltonian (%
\ref{DiracHamilSilic}) becomes block diagonal. For each spin $s_{z}=\pm 1$
and valley $\eta =\pm $, it describes a two-band system in the form, $H=%
\mathbf{\tau }\cdot \mathbf{d}$, where $d_{x}=\eta \hbar v_{\text{F}}k_{x}$, 
$d_{y}=\hbar v_{\text{F}}k_{y}$, $d_{z}=m_{\text{D}}$. The summation of the
Berry curvature is reduced to the Pontryagin index in the two-band system%
\cite{Qi}. They are determined uniquely by the Dirac mass and the valley
index. We explicitly have\cite{EzawaDiamag} 
\begin{equation}
\mathcal{C}_{s_{z}}^{\eta }={\frac{\eta }{2}}\text{sgn}(m_{\text{D}})
\end{equation}%
for the K$_{\eta }$ valley. The Chern and spin-Chern numbers are given by $%
\mathcal{C}=\sum_{\eta =\pm }(\mathcal{C}_{+}^{\eta }+\mathcal{C}_{-}^{\eta
})$ and $\mathcal{C}_{s}=\sum_{\eta =\pm }\frac{1}{2}(\mathcal{C}_{+}^{\eta
}-\mathcal{C}_{-}^{\eta })$, which are shown in the phase diagram (Fig.\ref%
{FigPhaseEA}). The phase boundaries are given by $m_{\text{D}}=0$ with (\ref%
{DiracMass}). A topological phase transition occurs when the sign of the
mass term changes.\textbf{\ }The topological numbers are $(C,C_{s})=(0,0)$
in the BI, $(0,1)$ in the QSHI, $(-2,0)$ in the P-QHI, and $(-1,\frac{1}{2})$
in the PS-QHI for $E>0$ and $\Omega >0$ in Fig.\ref{FigPhaseEA}. In all
these states the band gap is open, where the Fermi level is present, and
they are insulators. We have derived these results without the Rashba
interactions. They remain true when they are switched on adiabatically.

\begin{figure}[t]
\centerline{\includegraphics[width=0.5\textwidth]{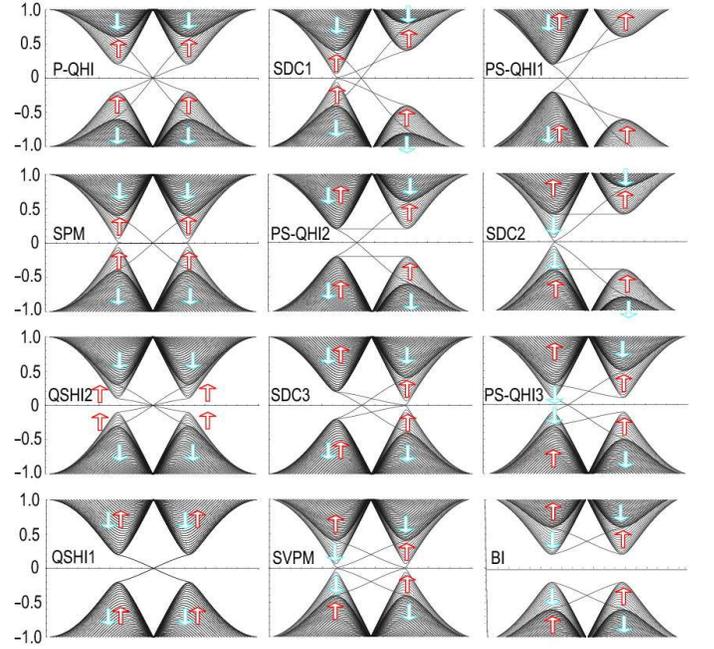}}
\caption{(Color online) The photon-dressed band structure of a silicene
nanoribbon at marked points in the phase diagram (Fig.\protect\ref%
{FigPhaseEA}). The vertical axis is the energy in uit of $t$, and the
horizontal axis is the momentum. We can clearly see the Dirac cones
representing the energy spectrum of the bulk. Lines connecting the two Dirac
cones are edge modes. The spin $s_{z}$ is practically a good quantum number,
which we have assigned to the Dirac cones. }
\label{FigPhotoRibbon}
\end{figure}

The Hall conductivity is given for each spin component by using the TKNN
formula\cite{TKNN}, $\sigma _{xy}^{s_{z}}=e^{2}/(2\pi \hbar )\sum_{\eta =\pm
}\mathcal{C}_{s_{z}}^{\eta }$. The charge-Hall and spin-Hall conductivities
are 
\begin{equation}
\sigma _{xy}=\sigma _{xy}^{\uparrow }+\sigma _{xy}^{\downarrow },\qquad
\sigma _{xy}^{s}=\sigma _{xy}^{\uparrow }-\sigma _{xy}^{\downarrow }.
\end{equation}%
They are equal to the Chern and spin-Chern numbers up to the normalization $%
e^{2}/2\pi \hbar $.

\begin{figure}[t]
\centerline{\includegraphics[width=0.4\textwidth]{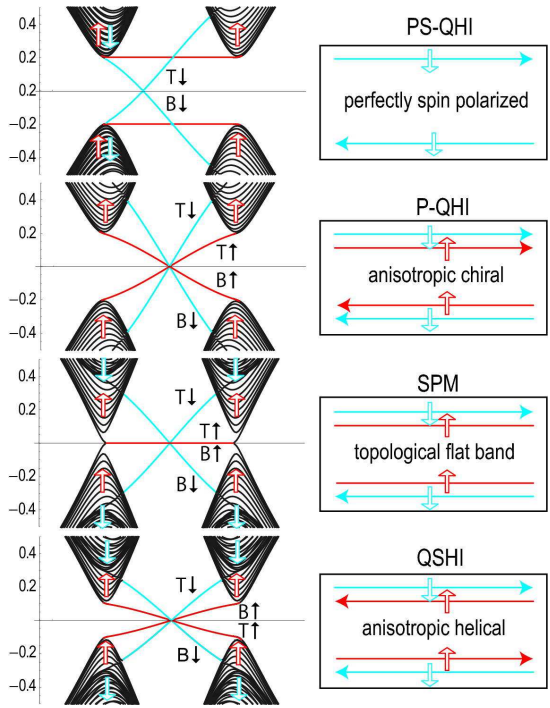}}
\caption{(Color online) Enlarged edge states of silicene nanoribbons for the
QSHI, SPM, P-QHI and PS-QHI. An anisotropic helical (chiral) current flows
in QSHI (P-QHI), where the velocities of up and down spins are different in
each edge. A topological flat band appears in SPM. Solid lines marked "T"
represent edge modes propagating on the top edge, while dotted lines marked
"B" represent edge modes propagating on the bottom edge. }
\label{FigEdgeIllust}
\end{figure}

\textbf{Photo-induced edge modes:} A further insight follows from the fact
that the commutator $[H_{-1},H_{+1}]$ is interpreted as the second-neighbor
hopping\cite{Kitagawa01B}. Hence, $\Delta H_{\text{eff}}$ is equivalent to
the Haldane model\cite{Haldane},%
\begin{equation}
\Delta H_{\text{eff}}=-i\eta \hbar v_{\text{F}}^{2}\mathcal{A}^{2}/(3\sqrt{3}%
\Omega )\sum_{\left\langle \!\left\langle i,j\right\rangle \!\right\rangle
\alpha \beta }\nu _{ij}c_{i\alpha }^{\dagger }c_{j\beta }.
\label{HaldaneTerm}
\end{equation}%
Based on this observation we have calculated the band structure of a
silicene nanoribbon with zigzag edges, which we give in Fig.\ref%
{FigPhotoRibbon} for typical points in the phase diagram (Fig.\ref%
{FigPhaseEA}). The edge mode changes between the chiral and helical states
by the topological phase transition.

\textbf{Conclusions:} We have discovered a class of new phases by applying
circular polorized light to silicene in the presence of electric field $%
E_{z} $, as summarized in the phase diagram (Fig.\ref{FigPhaseEA}) and in
the band structures of associated nanoribbons (Fig.\ref{FigPhotoRibbon}). We
summarize their typical features.

The P-QHI exhibits quantum Hall effects without magnetics field. They have
anisotropic chiral edge modes. It is remarkable that the velocities of up
and down spins in chiral edge states are different, as found by the
different slopes of the edge modes in Fig.\ref{FigEdgeIllust}. This is not
the case in graphene\cite{Oka09L}. Similarly, the velocities of up and down
spins in helical edge states are different in QSHI. The difference increases
as the intensity of light increases.

The SPM appears at the critical point between the QSHI and the P-QHI. It is
interesting to compare the state with the SVPM state appearing at the
critical point between the QSHI and the BI. Due to the identical spin
configuration at the K and K' points, there emerge spin polarized
topological flat bands in the SPM state (Fig.\ref{FigEdgeIllust}).

A particularly intriguing state is the SDC state emerging along the phase
boundary, where, e.g., the gap is open at the K point but closed at the K'
point. The spin is up polarized at K point and down polarized at K' point.
Thus the net spin is polarized in the SDC state. Hence, this state is
ferromagnet without magnetic field or exchange interactions. To create this
state we have broken the time-reversal and space-inversion symmetries. We
comment that there is no fermion doubling problem in the SDC state, because
the chiral symmetry is explicitly broken by the mass term.

In this paper we have studied the second order effect in the photo coupling $%
\mathcal{A}$. When the off-resonant condition $\hbar |\Omega |\gg t$ is
satisfied, there is no optical absorption. The lowest frequency is
determined by the band width $3t=4.8$eV$=10^{15}$Hz. Below this frequency,
the optical absorption occurs, which is the first order effect in $\mathcal{A%
}$. The peculiar optical selection rules and circular dichroism have already
been predicted in this regime\cite{EzawaOpti}.

\label{SecConclusion}

I am very much grateful to N. Nagaosa and T. Oka for many helpful
discussions on the subject. This work was supported in part by Grants-in-Aid
for Scientific Research from the Ministry of Education, Science, Sports and
Culture No. 22740196.

\end{document}